\documentclass[preprint,preprintnumbers,amsmath,amssymb,double-spaced]{revtex4}
\usepackage{graphicx}
\usepackage{epstopdf}
\usepackage{dcolumn}
\usepackage{bm}

\begin{document}

\title{Quantum speed-up transition in open system dynamics}

\author{Xiang Hao}
\email{xhao@phas.ubc.ca}

\affiliation{School of Mathematics and
Physics, Suzhou University of Science and Technology, Suzhou,
Jiangsu 215011, People's Republic of China \\}

\affiliation{Pacific Institute of Theoretical Physics, Department of Physics and Astronomy,
\\University of British Columbia, 6224 Agriculture Rd., Vancouver B.C., Canada V6T 1Z1.}

\author{Wenjiong Wu}

\affiliation{School of Mathematics and
Physics, Suzhou University of Science and Technology, Suzhou,
Jiangsu 215011, People's Republic of China}

\begin{abstract}

The rate of the trace distance is used to evaluate quantum speed-up for arbitrary mixed states. Compared with some present methods, the approach based on trace distance can provide an optimal bound to the speed of the evolution. The dynamical transition from no speed-up region to speed-up region takes on in the spontaneous decay of an two-level atom with detuning. The evolution is characteristic of the alternating behavior between quantum speed-up and speed-down in the strong system-reservoir coupling regime. Under the off-resonance condition, the dynamical evolution can be accelerated for short previous times and then decelerated to a normal process either in the weak or strong coupling regime. From the time-energy uncertainty relation, we demonstrate that the potential capacity for quantum speed-up evolution is closely related to the energy flow-back from the reservoir to the system. The negative decay rate for short time intervals leads to the speed-up process where the photons previously emitted by the atom are reabsorbed at a later time. The values of the spontaneous decay rate becomes positive after a long enough time, which results in the normal evolution with no speed-up potential.

\vspace{1.6cm} Keywords: quantum speed-up, dynamical transition, quantum decoherence, spontaneous decay


\end{abstract}

\maketitle

\section{Introduction}

Time is a valuable resource for the realization of fast information-processing devices. In the fields of quantum computation and quantum communication, much more attentions have been paid to quantum speed limit ($\mathrm{QSL}$)\cite{Giovanetti11,Margolus98,Lloyd00,Taddei13,Campo13,Deffner13,Fung14,Khalil15} for systems of interest in the recent years. The time-energy uncertainty relation\cite{Anandan1990,Vaidman1992,Giovannetti03,Jones10,Zwierz12,Pfeifer1995} gives constraints on the speed of quantum evolution. The $\mathrm{QSL}$ time can provide a minimal value of the time duration for the evolution of an initial state which jumps to a target state with certainty. The initial state can be distinguishable from the evolved target state. The $\mathrm{QSL}$ time is very useful for the estimation of the maximal rate of quantum information transferring \cite{Bekenstein81} and of quantum gate operation \cite{Lloyd00}. Besides the above applications, the concept of $\mathrm{QSL}$ plays a key role in quantum metrology \cite{Giovanetti11} and quantum optimal control algorithms \cite{Caneva09,Hegerfeldt13}.

For closed system, the evolution of states is unitary. The $\mathrm{QSL}$ depends on the Hamiltonian of systems $H$. In this case, the $\mathrm{QSL}$ time is referred to as the passage time during which an initial state can evolve to one orthogonal state. At present, the expression of the $\mathrm{QSL}$ time for closed systems can be written as the maximal value of Mandelstam-Tamm bound ($\mathrm{MT}$)\cite{Mandelstam1945} and Margolus-Levitin bound ($\mathrm{ML}$)\cite{Margolus98},  i.e., $\max\{ \pi\hbar/(2\Delta E),\pi\hbar/(2E)\}$. The $\mathrm{QSL}$ time is determined by the variance of the energy $\Delta E=\sqrt{\langle H^2 \rangle-\langle H \rangle^2}$ and the mean value of the energy $E=\langle H \rangle -E_0$ where $E_0$ is the initial energy. In fact, the system is unavoidably subject to the impacts of the environment due to the system-environment coupling. The dynamics of the open system is governed by a general non-unitary quantum evolutions \cite{Breuer02}. We can choose a variety of so-called discrimination measurement for two non-orthogonal states to obtain the quantum speed limit time, $\tau_{\mathrm{QSL}}$. So far, some methods for quantifying $\tau_{\mathrm{QSL}}$ have been put forward, including fidelity approach\cite{Campo13, Zhang14}, geometric Bures angle\cite{Taddei13,Deffner13} estimation and quantum Fisher information\cite{Frowis12} evaluation. In quantum information theory, the use of trace distance \cite{Nielsen00} is convenient to discriminate any two quantum states. Therefore, it is valuable to determine a lower bound on the minimal time duration in open system dynamics by means of trace distance method.

As we all know, the spontaneous decay of a two-level system is thought of as one typical kind of quantum decoherence dynamics. The decay process on resonance has been studied in the previous works \cite{Deffner13,Zhang14}. However, when a classic field is applied to the system, such quantum control often results in the off-resonance decay \cite{Solano03, Liu06}. Equivalently, the center frequency of the reservoir spectral density is detuned by an amount against the energy transition frequency of the system \cite{Breuer02}. Therefore, it is necessary to study the $\mathrm{QSL}$ time in the general spontaneous decay with detuning effects. Some important works have proven that the quantum speed-up process is connected with the non-Markovian effects \cite{Laine12,Ma14,WMZhang12,Fanchini14} of the open dynamics in the strong-coupling regime. That is, the non-Markovianity of the process can induce the decrease of the $\mathrm{QSL}$ time \cite{Xu14,Zhang15}. Besides it, the entanglement for open multi-particle systems can contribute to quantum acceleration \cite{Liu15}. We expect to find out other factors which initiate the speed-up evolution in the weak-coupling regime. The intuition tells us more work or energy must be needed to achieve the faster information processing. This reasonable assumption is based on the time-energy uncertainty relation. This principle in quantum mechanics motives us to think about one possible quantum speed-up mechanism for open systems dynamics, from the point of view of the energy flow.

The paper is organized as follows. In Sec. II, arbitrary quantum non-unitary evolution is considered to obtain the expression of $\mathrm{QSL}$ time based on trace distance. It is found out that the $\mathrm{QSL}$ time estimation can be applicable for arbitrary initial states in open systems dynamics. In Sec. III, the general spontaneous decay of a two-level system with detuning will be introduced. The time-dependent decay rate can be used to describe the energy flow between the system and the reservoir. The impacts of the system-coupling strength and detuning effects on $\mathrm{QSL}$ time are studied in Sec. IV. It is shown that there is the dynamical transition from the quantum speed-up process to no speed-up process. The physical explanation for this dynamical transition is given in detail. In Sec. V, a discussion concludes the paper.

\section{Approach for $\mathrm{QSL}$ time}

In order to attain an optimal bound on the $\mathrm{QSL}$ time, we need to choose appropriate methods for discriminating any two quantum states. In \cite{Deffner13}, one geometric method on the basis of Bures angle was introduced to estimate the bound on $\mathrm{QSL}$. This method can provide a tight bound to the speed of the evolution under a certain condition. In our framework, we put forward another approach on the basis of trace distance. According to \cite{Nielsen00}, the trace distance $D_{\rho}(t,0)$ between an arbitrary initial state $\rho_{0}$ and evolved state $\rho_{t}$ after a certain time $t$ is defined as,
\begin{equation}
\label{eq:(1)}
D_{\rho}(t,0) \;=\; 1-\frac{1}{4}\left \|\rho_{t}-\rho_{0}\right \|_1^2.
\end{equation}
Here the norm is calculated by $\left \| \mathrm{A} \right \|_1=\mathrm{Tr} \left \{ \sqrt{\mathrm{A}^{\dag} \mathrm{A} } \right \}$ \cite{Bhatia1997}. We make use of a general time-dependent non-unitary evolution to describe the open system dynamics as
\begin{equation}
\label{eq:(2)}
\dot{\rho}_t \;=\; L_t(\rho_t),
\end{equation}
where $L_t$ are trace class super-operators in a complex Banach space. In the following analysis, the condition of $|| L^{\dag}_t||_p=|| L_t||_p$ is considered. The Schatten $p$ norm $||L_t||_p=\left[ \sum_{i}\lambda_i^p \right]^{1/p}$ where $\lambda_i$ are the singular values of the operator $L_t(\rho_t)$ in the descending sequence, i.e., $\lambda_1$ is the maximal singular value. The time derivative of the trace distance is used to study the dynamical velocity with which the density operator of the open system evolves. The rate of trace distance satisfies the following inequality as,
\begin{equation}
\label{eq:(3)}
\left| \frac{\mathrm{d} D_{\rho}}{\mathrm{d} t} \right| \; \leq \; \frac{1}{2}\left \|\rho_{t}-\rho_{0}\right \|_1 \; \left|\mathrm{Tr}\left \{ \dfrac{\rho_{t}-\rho_{0}}{\sqrt{(\rho_{t}-\rho_{0})^2}}L_t(\rho_t) \right\} \right|.
\end{equation}
Here the triangle inequality $|a\pm b|\leqslant|a|+|b|$ and trace relation of $\mathrm{Tr}\left \{ AB \right\}=\mathrm{Tr}\left \{ BA \right\}$ are used to obtain the above inequality.

Through von Neumann trace inequality\cite{Grigorieff1991}, we have the result of  $|\mathrm{Tr}\{\frac{\rho_t-\rho_0}{\sqrt{(\rho_t-\rho_0)^2}}L_t(\rho_t)\}|\leq \sum_{i=1}^n \sigma_i \; \lambda_i$.
$\{\sigma_i=1, \, 1, \, \dots, \, 1 \}$ are the singular values of the matrix $\frac{\rho_t-\rho_0}{\sqrt{(\rho_t-\rho_0)^2}}$ and $n$ is the Hilbert-space dimension of the system. Because of $\sum_i \lambda_1 =n|| L_t||_{\infty} \geq \sum_i \lambda_i=||L_t||_1$, the absolute value of the rate of the trace distance is bounded by
\begin{align}
\label{eq:(4)}
|\frac{\mathrm{d} D_{\rho}}{\mathrm{d}t}| &\leq \; \frac{1}{2}\left \|\rho_t-\rho_0\right \|_{1} \; \left \|L_t(\rho_t)\right \|_{1} \nonumber \\
&\leq \; \frac{1}{2}\left \|\rho_t-\rho_0\right \|_1 \; n\left \|L_t(\rho_t)\right \|_{\infty}.
\end{align}
When the actual driving time for open systems are chosen to be $t=\tau_D$, the time duration has such constraints as
\begin{equation}
\label{eq:(5)}
\tau_D \; \geq \; \max \{\frac{1}{\Lambda_{\tau_D}^1},\frac{1}{\Lambda_{\tau_D}^{\infty}} \} \; 2|1-D_{\rho}(\tau_D,0)|,
\end{equation}
where $\Lambda_{\tau_D}^1=\frac{1}{\tau_D}\int_{0}^{\tau_D} \left \|\rho_t-\rho_0\right \|_1 \; \left \|L_t(\rho_t)\right \|_1\, \mathrm{d}t$ and $\Lambda_{\tau_D}^{\infty}=\frac{n}{\tau_D}\int_{0}^{\tau_D} \left \|\rho_t-\rho_0\right \|_1 \; \left \|L_t(\rho_t)\right \|_{\infty}\, \mathrm{d}t$.
The above formula is reduced to the $\mathrm{ML}$ bound for a closed system. In this case, a couple of orthogonal states is chosen as the initial and target state. The unitary evolution of the closed system is determined by $\dot{\rho}_t=\frac{1}{i\hbar}[H_t,\rho_t]$ where $H_t$ is the time-dependent Hamiltonian. Therefore, $\left \|[H_t,\rho_t]\right \|_1=\left \|H_t\rho_t-\rho_tH_t\right \|_1 \leq \left \|H_t\rho_t\right \|_1+\left \|\rho_tH_t\right \|_1=2\left \|H_t\rho_t\right \|_1=2 \left\langle H_t\right\rangle$, and the trace distance between any two orthogonal states satisfies that $||\rho_{\tau_D}-\rho_0||_1=2$. When we substitute these equations into Eq.(5), $\mathrm{ML}$-type bound for a closed system with $n=2$ is obtained as $\tau_D\geq \hbar/2E_{\tau_D}$ where $E_{\tau_D}=\frac {1}{\tau_D}\int_{0}^{\tau_D} \left\langle H_t\right\rangle \, \mathrm{d}t$ is the mean energy during the driving time interval.

In another way, the Cauchy-Schwarz inequality \cite{Bhatia1997} $|\mathrm{Tr}\{A_1A_2\}|\leq \sqrt{\mathrm{Tr}\{A_1A_1^\dag\}\; \mathrm{Tr}\{A_2A_2^\dag}\}$ can be used to simply the Eq. (3). We can obtain $|\mathrm{Tr}\{\frac{\rho_t-\rho_0}{\sqrt{(\rho_t-\rho_0)^2}}L_t(\rho_t)\}|\leq \sqrt{n}\cdot
\left \|L_t(\rho_t)\right \|_2$. Consequently, the time duration is also bounded by
\begin{equation}
\label{eq:(6)}
\tau_D \; \geq \; 2|1-D_{\rho}(\tau_D,0)|\; \frac{1}{\Lambda_{\tau_D}^2},
\end{equation}
where $\Lambda_{\tau_D}^2=\frac{\sqrt{n}}{\tau_D}\int_{0}^{\tau_D} \left \|\rho_t-\rho_0\right \|_1 \; \left \|L_t(\rho_t)\right \|_2\, dt$.
Similarly, the above bound can also be reduced to be the $\mathrm{MT}$ type bound for the closed system. In this case, $\left \|H_t\rho_t-\rho_tH_t\right \|_2^2=2 \mathrm{Tr}\{H_t^2\rho_t^2\}-2\mathrm{Tr}\{(H_t\rho_t)^2\} \leq 2 \mathrm{Tr}\{H_t^2\rho_t \}-2(\mathrm{Tr}\{H_t\rho_t\})^2$. Thus, the $\mathrm{MT}$ type bound for a closed system with $n=2$ is acquired by $\tau_D\geqslant\frac{\hbar}{2\Delta E_{\tau_D}}$ where $E_{\tau_D}=\frac {1}{\tau_D}\int_0^{\tau_D} \sqrt{\langle H_t^2 \rangle-\langle H_t \rangle^2}$ is the variance of the energy for the driving time. Therefore, it is reasonable to adopt the trace distance as one approach to describe $\tau_{QSL}$ for the open system dynamics.

From what has been discussed above, the bound on the quantum speed limit is given as,
\begin{equation}
\label{eq:(7)}
\tau_{QSL} \;=\; \max \{\frac{1}{\Lambda_{\tau_D}^1},\frac{1}{\Lambda_{\tau_D}^2}, \frac{1}{\Lambda_{\tau_D}^{\infty}}\}\; 2|1-D_{\rho}(\tau_D,0)|.
\end{equation}
It is shown that the above equation for the $\mathrm{QSL}$ time is applicable to arbitrary initial state, i.e., $\rho_0$ is a general mixed state. This estimation of the $\mathrm{QSL}$ time is reduced to the results of the closed system dynamics, and also consistent with the unified time bound on $\mathrm{QSL}$ in the previous works. Therefore, the validity of our result is confirmed.

\section{Spontaneous decay model with detuning}

As an specific example we treat the decoherence model of a two-level atom which decays spontaneously into a field vacuum of a structural reservoir at zero temperature. The general off-resonance case is taken into account. In the following study, we use an approach based on trace distance to estimate the speed-up process of the open system. This system coupled to the zero-temperature structural environment with a detuning spectral density can be described by the Hamiltonian,
\begin{equation}
\label{eq:(8)}
H \;=\; H_{A}+H_{R}+H_{I} \;=\;H_{0}+H_{I},
\end{equation}
where the respective Hamiltonians of the atom $A$ and reservoir $R$, the atom-reservoir coupling Hamiltonian are given by
\begin{align}
\label{eqs:(9-10)}
H_{0}&= \; \frac {\omega_0}2 \sigma_{z}+\sum_{k} \omega_{k} b^{\dag}_{k} b_{k}, \\
H_{I}&=\; \sigma_{+}\otimes B+\sigma_{-} \otimes B^{\dag}  \; \mbox{with} \; B=\sum_k \; g_k b_k.
\end{align}
The rotating wave approximation is applied here. The transition frequency of the atom is denoted by $\omega_0$, and $\sigma_{\pm}$ are the raising and lowering operators which have $\sigma_{+}|g\rangle_A=|e\rangle_A$ and $\sigma_{-}|e\rangle_A=|g\rangle_A$. Here $|e(g)\rangle_A$ represents the excited(ground) state for the system. The index $k$ labels all field modes of the reservoir with different frequencies $\omega_k$, creation and annihilation field operators $b^{\dag}_{k}, \; b_{k}$. And $g_k$ represents the coupling strength between the atom and reservoir. The Hamiltonian of the model can be physically realized by the two-level system coupled to electromagnetic fields in a leaky cavity, which is illustrated in Figure 1. This is an exactly solvable model, which is referred to as a reasonable one for the evaluation of the speedup potential capacity. Because of the commutation $[H, \sigma_z+\sum_k b^{\dag}_k b_k]=0$, the quality of the number operator is conserved. When the initial state is chosen as $|\Psi(0)\rangle=|e\rangle_A \otimes |0\rangle_R$, we can express the time-dependent total state in the subspace spanned by $\{|e\rangle_A|0\rangle_R, |g\rangle_A|k\rangle_R\}$ as,
\begin{equation}
\label{eq:(11)}
|\Psi(t)\rangle \;=\; C(t)|e\rangle_A|0\rangle_R+ \sum_k C_k(t)|g\rangle_A|k\rangle_R,
\end{equation}
where $|0\rangle_R$ is the vacuum field state, and $|k\rangle_R=b^{\dag}_k|0\rangle_R$ denotes the state with one photon in mode $k$. In the interaction picture, the above time-dependent state obeys the Shr\"{o}dinger equation with the interaction Hamiltonian $H_I(t)=\mathrm{e}^{iH_0t}H_I \mathrm{e}^{-iH_0t}$. The time-dependent coefficient $C(t)$ is derived from a series of differential equations,
\begin{align}
\label{eq:(12)}
\dot{C}(t)&= \; -i\sum_k g_k \exp[i(\omega_0-\omega_k)t]C_k(t) \nonumber \\
\dot{C}_k(t) &= \; -ig^{*}_k \exp[-i(\omega_0-\omega_k)t]C(t)
\end{align}
We can firstly solve the second equation and then insert it into the first equation to get a closed equation $\dot{C}(t)=-\int_{0}^{t} f(t-t_1)C(t)\, \mathrm{d} t_1$ where the kernel $f(t-t_1)$ is given by the correlation function $f(t-t_1)=\sum_k |g_k|^2 \mathrm{Tr}_{R} \{b_k b^{\dag}_k |0\rangle_R \langle 0| \}\exp[i(\omega_0-\omega)(t-t_1)]$. The atom-field couplings can be expressed by the detuning Lorentzian spectral density $J(\omega)$,
\begin{equation}
\label{eq:(13)}
J(\omega) \;=\; \frac{1}{2}\frac{\gamma_0\lambda^2}{(\omega_0-\Delta-\omega)^2+\lambda^2}.
\end{equation}
The parameter $\gamma_0$ is the coupling strength, $\lambda$ the spectral width. $\Delta$ is the amount for the centre frequency $\omega_c$
of the cavity detuned against the atomic transition frequency $\omega_0$. When
$\gamma_0<\frac{1}{2}\lambda$, the dynamics is in the weak coupling regime. The case of $\gamma_0>\frac{1}{2}\lambda$ denotes the evolution process  in a strong coupling regime. Consequently, the correlation function $f(t-t_1)$ is expressed by
\begin{equation}
\label{eq:(14)}
f(t-t_1) \;=\; \frac{1}{2}\gamma_0\lambda \;\exp[-(\lambda-i\Delta)(t-t_1)].
\end{equation}
The time-dependent reduced density matrix of the two-level atom $A$ is expressed by \cite{Garraway97,Breuer02,Maniscalco06}
\begin{equation}
\label{eq:(15)}
\rho_t \;=\; \begin{pmatrix}
          \rho_{11}(0)|C(t)|^2 & \rho_{10}(0)C(t) \\
          \rho_{01}(0)C^\ast(t) & \rho_{11}(0)(1-|C(t)|^2)
        \end{pmatrix}.
\end{equation}

For the description of the reduced dynamics of the atom, we can make use of the time-dependent quantum master equation in the form of
\begin{align}
\label{eq:(16)}
\frac {\mathrm{d}}{\mathrm{d}t}\rho_t \;=& L_t(\rho_t)=-\frac {i}{2} S(t) [\sigma_z,\rho_t] \nonumber\\ &
+ \gamma(t)(\sigma_-\rho_t\sigma_+ - \frac{1}{2}\sigma_+\sigma_-\rho_t - \frac{1}{2}\rho_t\sigma_+\sigma_-).
\end{align}
$S(t)=-2 \mathrm{Im} \left\{ \frac {\dot{C(t)}}{C(t)} \right\}$ represents the impacts of the Lamb shift, and $\gamma(t)=-2 \mathrm{Re} \left\{ \frac {\dot{C(t)}}{C(t)} \right\}$ is the time-dependent decay rate which is dominant in the dynamical process. After this correlation function is solved, the expression of $C(t)$ is written as,
\begin{equation}
\label{eq:(17)}
C(t) \;=\; e^{-\frac{1}{2}(\lambda-i\Delta)t}\left[\cosh\left(\frac{dt}{2}\right)+\frac{\lambda-i\Delta}{d}\sinh\left(\frac{dt}{2}\right)\right],
\end{equation}
where $d=\sqrt{(\lambda-i\Delta)^2-2\gamma_0\lambda}$. The positive values of $\gamma(t)$ leads to a positive contribution to the ground state population $\rho_{00}(t)$ while the negative values results in the increase of the excite state population. The latter one denotes the flow-back of the energy from the reservoir to the system.

\section{Quantum acceleration dynamical transition}

To effectively estimate the quantum speed-up process, we reasonably assume that one initial state and one target state can be distinguished by the trace distance after an actual driving time $\tau_D$. Different from the case of closed system dynamics, these two discriminate states are not orthogonal in the open system evolution. In the following analysis, we can find out that the minimal time duration for distinguishing the evolution states is determined by the $\tau_{QSL}$. The value of $\tau_{QSL}/\tau_D$ estimates the capacity of a quantum system to accelerate in dynamics.
The smaller $\tau_{QSL}/\tau_D$ is, the stronger this ability is. The speed-up region in the dynamics corresponds to the values $\frac {\tau_{QSL}}{\tau_D}<1$ while the values $\frac {\tau_{QSL}}{\tau_D}=1$ represent no speed-up process.

For the spontaneous decay model with detuning, the excited state $|e\rangle_A$ is selected as the initial state in the decoherence dynamics. That is, the initial condition at time $t=0$ is $\rho_{11}(0)=1, \rho_{10}(0)=\rho_{01}(0)=0$. We have $\Lambda_{\tau_D}^1=\Lambda_{\tau_D}^2<\Lambda_{\tau_D}^{\infty}$. In this case, the $\mathrm{ML}$ bound is as tight as the $\mathrm{MT}$ bound. The expression of $\tau_{QSL}/\tau_D$ is written as
\begin{equation}
\label{eq:(18)}
\frac{\tau_{QSL}}{\tau_D}\;=\; 2|1-D_{\tau_D}| \; \int_{0}^{\tau_D} \left \|\rho_t-\rho_0\right \|_1 \; \left \|L_t(\rho_t)\right \|_1 \, \mathrm{d}t.
\end{equation}

To further examine the usefulness of our approach, we compare the numerical results in accordance with the two different methods. It is shown that the speed-up region attained by the trace distance is almost equivalent to that by the Bures angle in Figure 2. This result means that our approach based on trace distance is as valid as the method of \cite{Deffner13}. Under the resonance condition, the method based on Bures angle can provide a bound which is tighter than that obtained by the trace distance. However, in the off-resonance case, we can see that a sharp bound obtained by the trace distance is optimal in contrast with the bound by Bures angle.

Figure 3 shows the ratio $\tau_{QSL}/\tau_D$ as a function of the coupling strength $\gamma_0$ and detuning $\Delta$. The actual driving time is chosen to be $\tau_D=0.2$. The width of the spectral density $\lambda=50$. In the very strong coupling regime, there always exists the potential of capacity for the quantum speed-up dynamics. Under the conditions of weak couplings and small detuning amount, $\tau_{QSL}/\tau_D=1$ demonstrates a plateau, which denotes no speed-up process. However, it is interesting that the large detuing amount can induce the accelerating evolution in the weak-coupling regime. Such quantum speed-up transition is clearly shown in Figure 4. It is seen that there are the speed-up region for $\frac {\tau_{QSL}}{\tau_D}<1$ and no speed-up region for $\frac {\tau_{QSL}}{\tau_D}=1$. The dynamical transition from no speed-up to speed-up occurs in the weak-coupling case. Specially, the increase of the detuning amount leads to the alternating transitions in the strong-coupling case.

To explain the physical mechanism of such quantum speed-up transition, we investigate the property of the $\mathrm{QSL}$ time when the initial states are determined by the arbitrary evolved states $\rho_{\tau}$. In what follows, the integral interval of Eq.(18) is changed from $t=\tau$ to $t=\tau+\tau_D$ \cite{Zhang14}. It is also assumed that the atom spontaneously decay from the excited state. The reduced density matrices for the evolved states at an arbitrary time $\tau$ is obtained by
\begin{equation}
\label{eq:(19)}
\rho_{\tau}\;=\; \begin{pmatrix}
          P_{\tau} & 0 \\
          0 & 1-P_{\tau}
        \end{pmatrix},
\end{equation}
where the parameter $P_{\tau}=|C(\tau)|^2$ denotes the population of the excited state. The rate of the population $\dot{P_t}>0$ represents the increase of the excited-state population which results from the energy flowing back from the reservoir to the atom. That is, the photons previously emitted by the atom are reabsorbed at a later time. If $\dot{P_t}<0$, the decrease of the population describes the energy-decaying process of the atom into the reservoir. When the evolved states are selected as the initial states, the ratio $\tau_{QSL}/\tau_D$ becomes
\begin{equation}
\label{eq:(20)}
\frac{\tau_{QSL}}{\tau_D}\;=\;\dfrac{(P_{\tau+\tau_D}-P_\tau)^2}{2\int_{\tau}^{\tau+\tau_D}
|(P_t-P_\tau)\dot{P_t}| \mathrm{d}t}.
\end{equation}
It is clearly seen from Figure 5(a) that the values of $\tau_{QSL}/\tau_D$ keep one in the weak-coupling regime with no detuning effects. And the values $\tau_{QSL}/\tau_D$ oscillate periodically in the strong-coupling regime with change of the evolved states. This is the reason that the population of the excited state always declines with time $\tau$, i.e., $\dot{P_t}<0$. In this case, the calculation result of the $\mathrm{QSL}$ time is simplified as $\tau_{QSL}/\tau_D=1$ which means that no speed-up process happens. However, the strong atom-reservoir couplings result in the oscillations of the population $P_{\tau}$. It is found out that the change of the population is decided by the time-dependent decay rate. The decay rate in the on-resonance case is given by,
\begin{equation}
\label{eq:(21)}
\gamma_{\mathrm{on}}(t) \;=\; \dfrac {2\gamma_0 \lambda\; \sinh(d_0\; t/2)} {d_0 \; \cosh(d_0\; t/2)+\lambda\; \sinh(d_0\; t/2)},
\end{equation}
where $d_0=\sqrt{\lambda^2-2\gamma_0 \lambda}$. Figure 5(b) shows the decay rate varies with the evolved time. The positive values of $\gamma_{\mathrm{on}}(t)$ denote the decrease of the population which cannot induce the quantum speed-up process. On the other hand, the negative values of $\gamma_{\mathrm{on}}(t)$ give rise to the increase o the population which contributes to the occurrence of quantum acceleration. This phenomena is consistent with the intuition that the more energy that the system keeps is helpful for the fast evolution.

The physical explanation for the quantum speed-up transition is also applicable in the general off-resonance case. Figure $6(a)$ illustrates the change of the ratio $\tau_{QSL}/\tau_D$ with the evolved time $\tau$ under the large-detuning condition. Compared with the on-resonance results, Figure $6(a)$ shows that the prominent detuning effect can switch on the potential capacity for quantum acceleration in the weak-coupling regime. On the circumstance of the equivalent detuning effect, the acceleration phenomenon for the strong-coupling regime is more obvious than that for the weak-coupling regime. It is also shown that the quantum evolution previously experiences the quantum speed-up process, and then slows down to the normal process. This result is distinct from that of the on-resonance decay. We can also give a reasonable explanation based on the energy flow. The off-resonance decay rate $\gamma_{\mathrm{off}}(t)=-2 \mathrm{Re} \left\{ \frac {\dot{C(t)}}{C(t)} \right\}$ can be obtained by the Eq. (17). Figure 6(b) is plotted to describe the change of the decay rate with the evolved time. In both weak-coupling and strong-coupling regime, the oscillations of the decay rate can be suppressed with the time. The negative values of $\gamma(t)$ takes on during the previous short time interval. From it, we know that the increase of the excited-state population can be maintained for a short time interval. This physical description can be mathematically demonstrated by the expression of the decay rate. After a long enough time, the decay rate reduces to the Markovian limit,
\begin{equation}
\label{eq:(22)}
\gamma_{M}\;=\; \frac {\gamma_0 \lambda^2}{\lambda^2+\Delta^2},
\end{equation}
which is always positive. Therefore, the previous acceleration process gradually decelerates to the normal evolution without any speed-up potential capacity. In a word, the quantum acceleration is mainly determined by the energy flow-back from the reservoir to the open system. Such dynamical transition is dependent on the mechanism of the energy flow.

\section{Conclusions}
The general quantum speed limit time for open system dynamics is derived on the basis of the trace distance. The unified expression is similar to the previous results in the refs. The general off-resonance spontaneous decay model is considered. We can demonstrate that the large-detuning effects and strong system-reservoir couplings can manifestly induce the acceleration dynamics. Compared with the on-resonance decay dynamics, the quantum speed-up process starts for the previous short time intervals, and then gradually undergoes one normal evolution with no acceleration. Because we take into account the spontaneous decay of an excited atom, the increase of the excited-state population denotes the occurrence of the energy flowing from the reservoir to the atom. That is, the atoms reabsorbs the photons previously emitted. In this photon-reabsorption process, the atom can maintain more energy for a short time interval. According to the time-energy uncertainty relation, the more energy the system keeps, the fast evolution the system experiences.

\newpage

\begin{figure}
  \centering
  \includegraphics[width=0.7\textwidth]{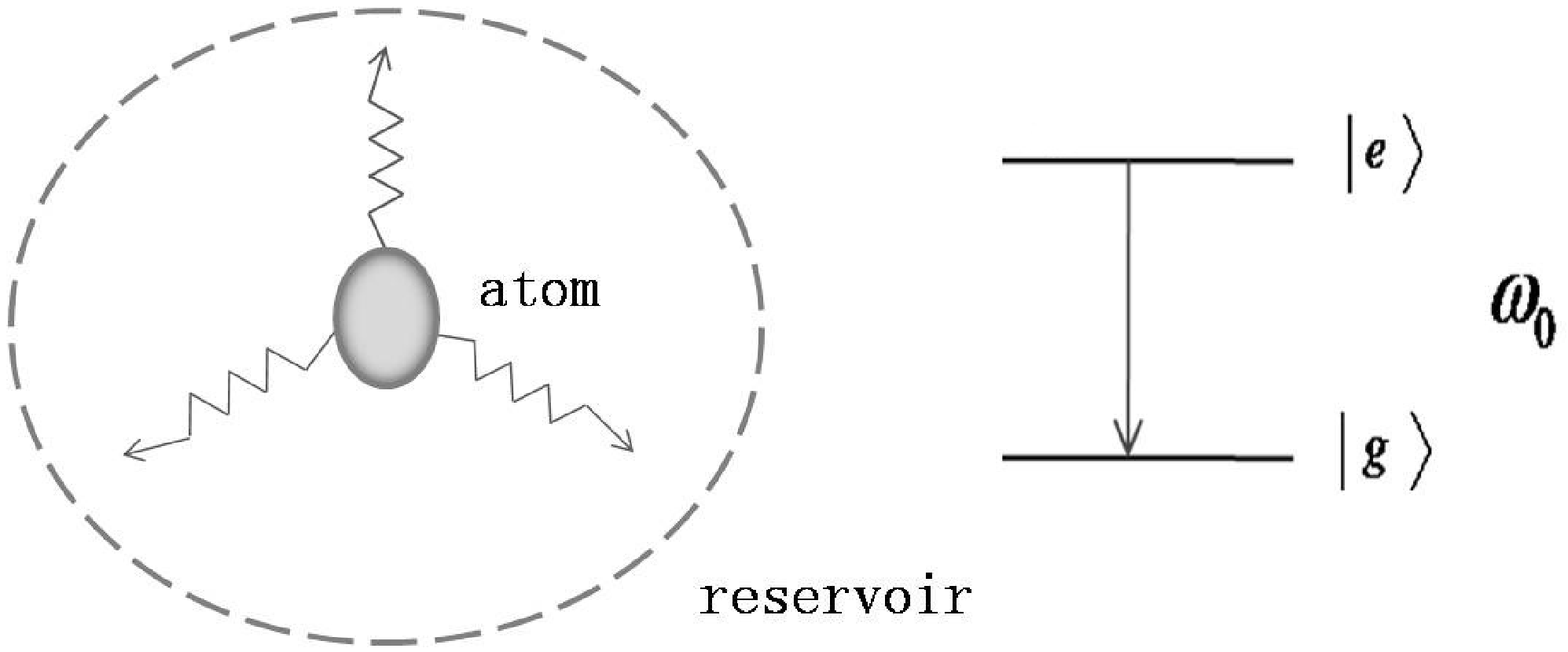}\\
  \caption{The physical model of the spontaneous decay of the two-level atom is illustrated. $\omega_0$ denotes the transition frequency between the excited state and ground state.}\label{1}
\end{figure}

\begin{figure}
  \centering
  \includegraphics[width=0.7\textwidth]{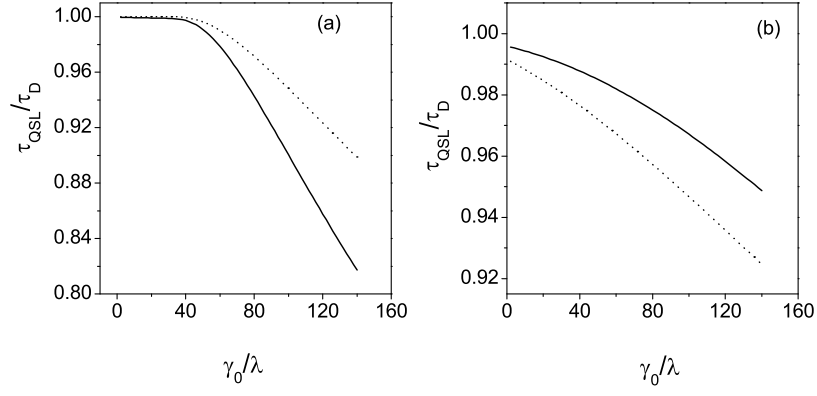}\\
  \caption{The numerical results of $\tau_{QSL}/\tau_D$ based on two different measures are compared. $(a)$ : The on-resonance condition of $\Delta=0$ is considered; $(b)$ : The off-resonance condition of $\Delta=4\lambda$ is assumed. The width of the spectral density is $\lambda=50$. The solid line represents the results from the trace distance and dashed line denotes those from Bures angles.}\label{2}
\end{figure}

\begin{figure}
  \centering
  \includegraphics[width=0.7\textwidth]{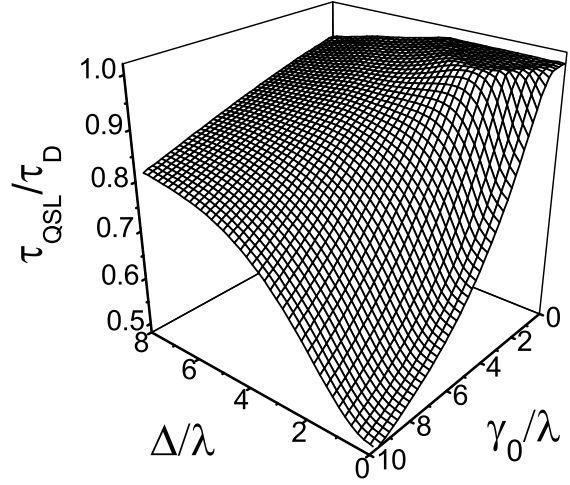}\\
  \caption{$\tau_{QSL}/\tau_D$ as a function of the coupling strength $\gamma_0$ and detuning parameter $\Delta$ is plotted when the initial state
  is the excited state. The width of the spectral density is $\lambda=50$.}\label{3}
\end{figure}

\begin{figure}
  \centering
  \includegraphics[width=0.4\textwidth]{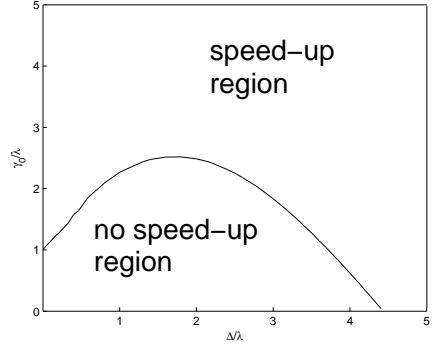}\\
  \caption{The transition between the speed-up region and no speed-up region is plotted with change of the coupling strength $\gamma_0$
  and detuning parameter $\Delta$. The width of the spectral density is $\lambda=50$.}\label{4}
\end{figure}

\begin{figure}
  \centering
  \includegraphics[width=0.4\textwidth]{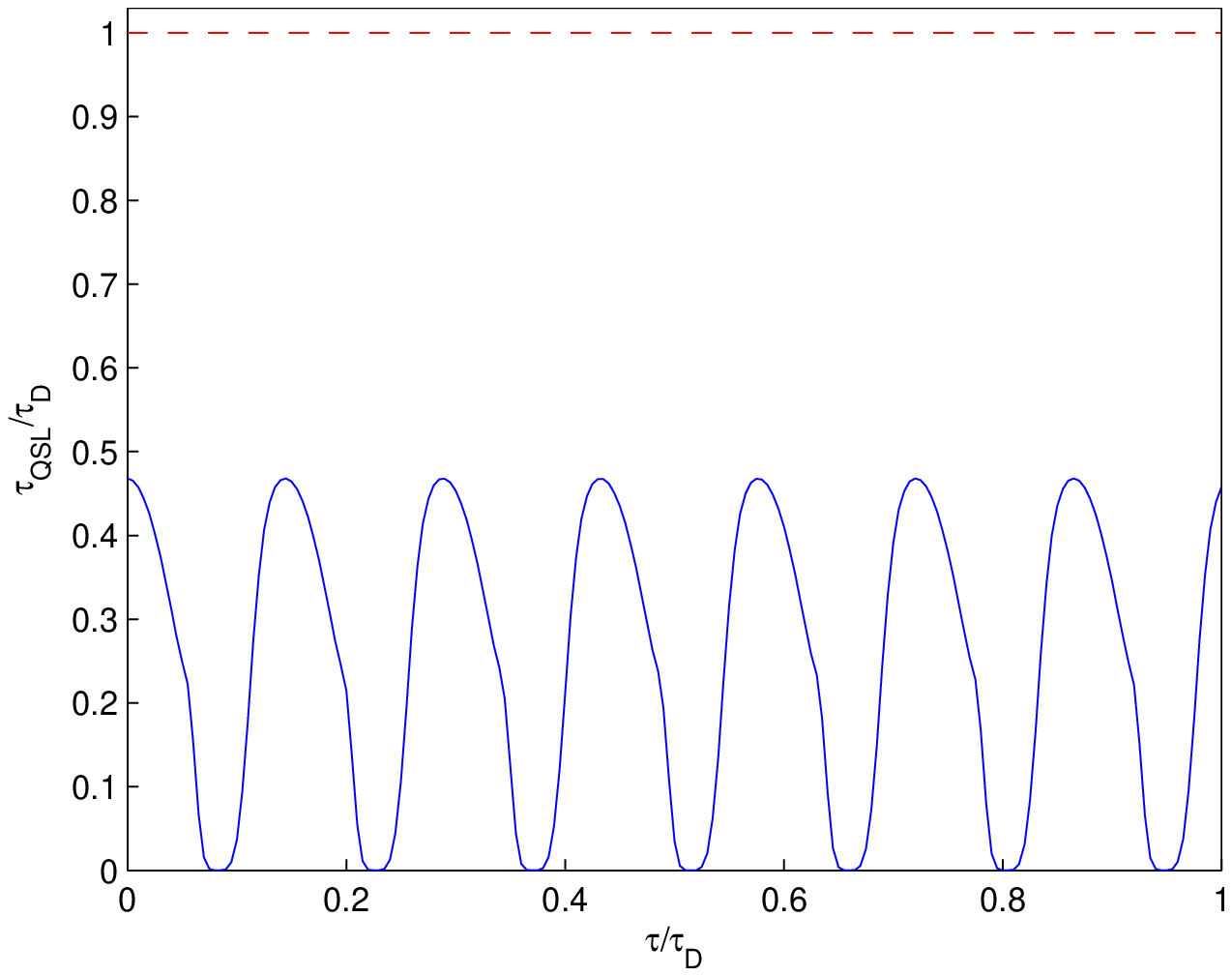}\\
  \includegraphics[width=0.4\textwidth]{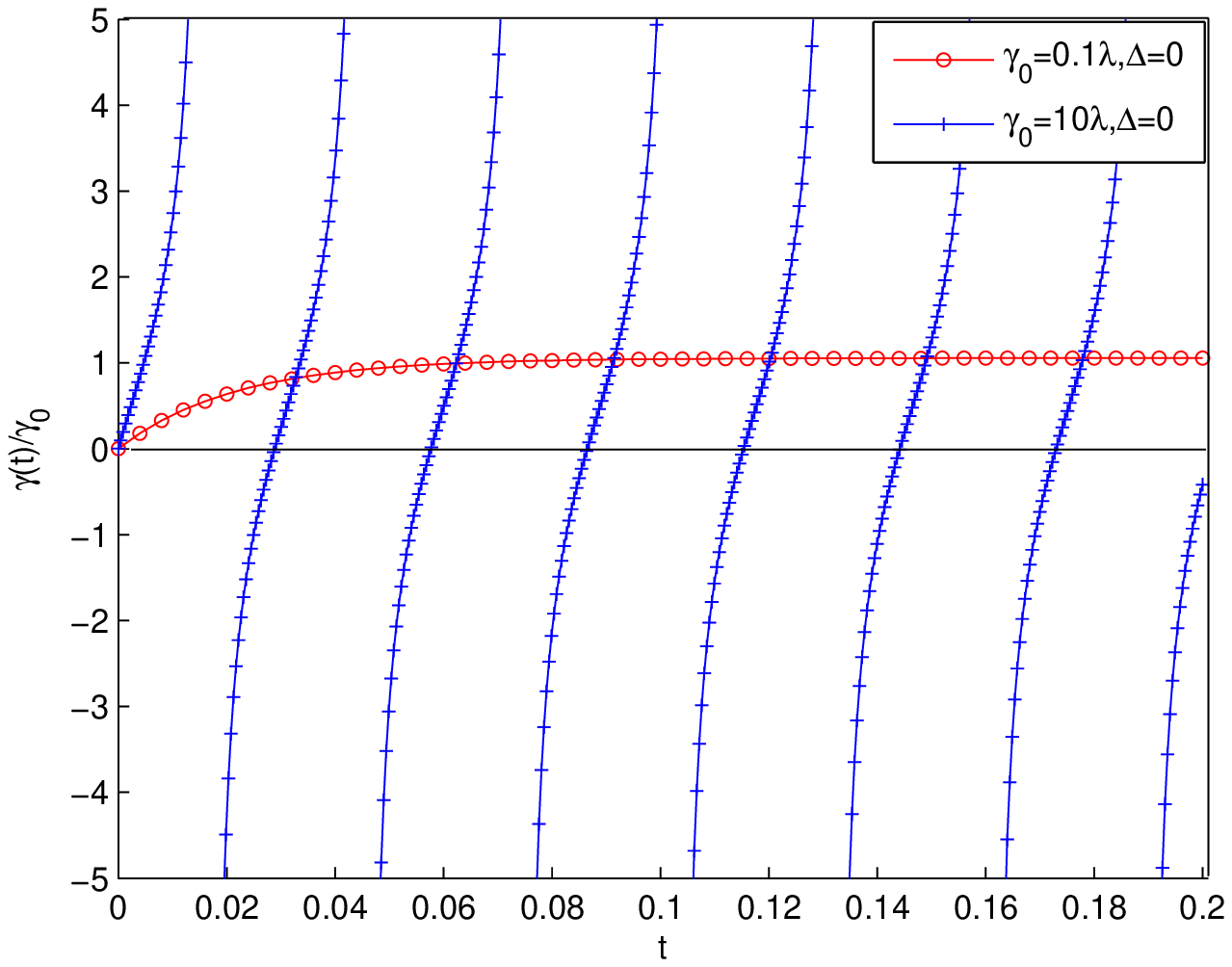}\\
  \caption{$(a)$ : $\tau_{QSL}/\tau_D$ is plotted as a function of the evolved time $\tau$ in the on-resonance case.
  The red dashed line represents the result of the weak-coupling regime, $\gamma_0=0.1\lambda, \; \Delta=0$.
  The blue solid line represents the result of the strong-coupling environment, $\gamma_0=10\lambda, \; \Delta=0$. $(b)$ : the on-resonance decay rate $\gamma(t)/\gamma_0$ changes with time. The red circles denote the condition of $\gamma_0=0.1\lambda, \; \Delta=0$. The blue pluses denote that of $\gamma_0=10\lambda, \; \Delta=0$.}\label{5}
\end{figure}

\begin{figure}
  \centering
  \includegraphics[width=0.4\textwidth]{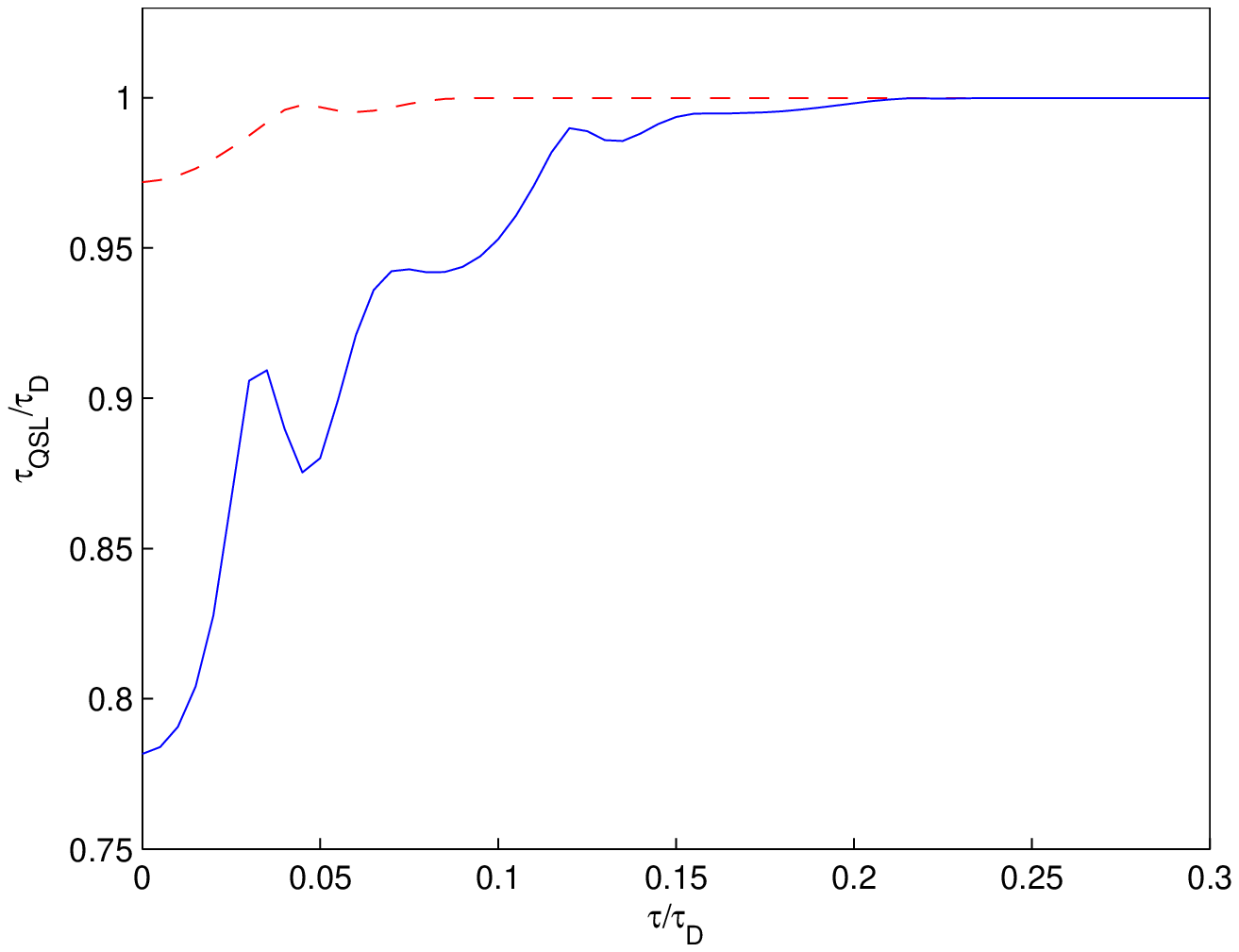}\\
  \includegraphics[width=0.4\textwidth]{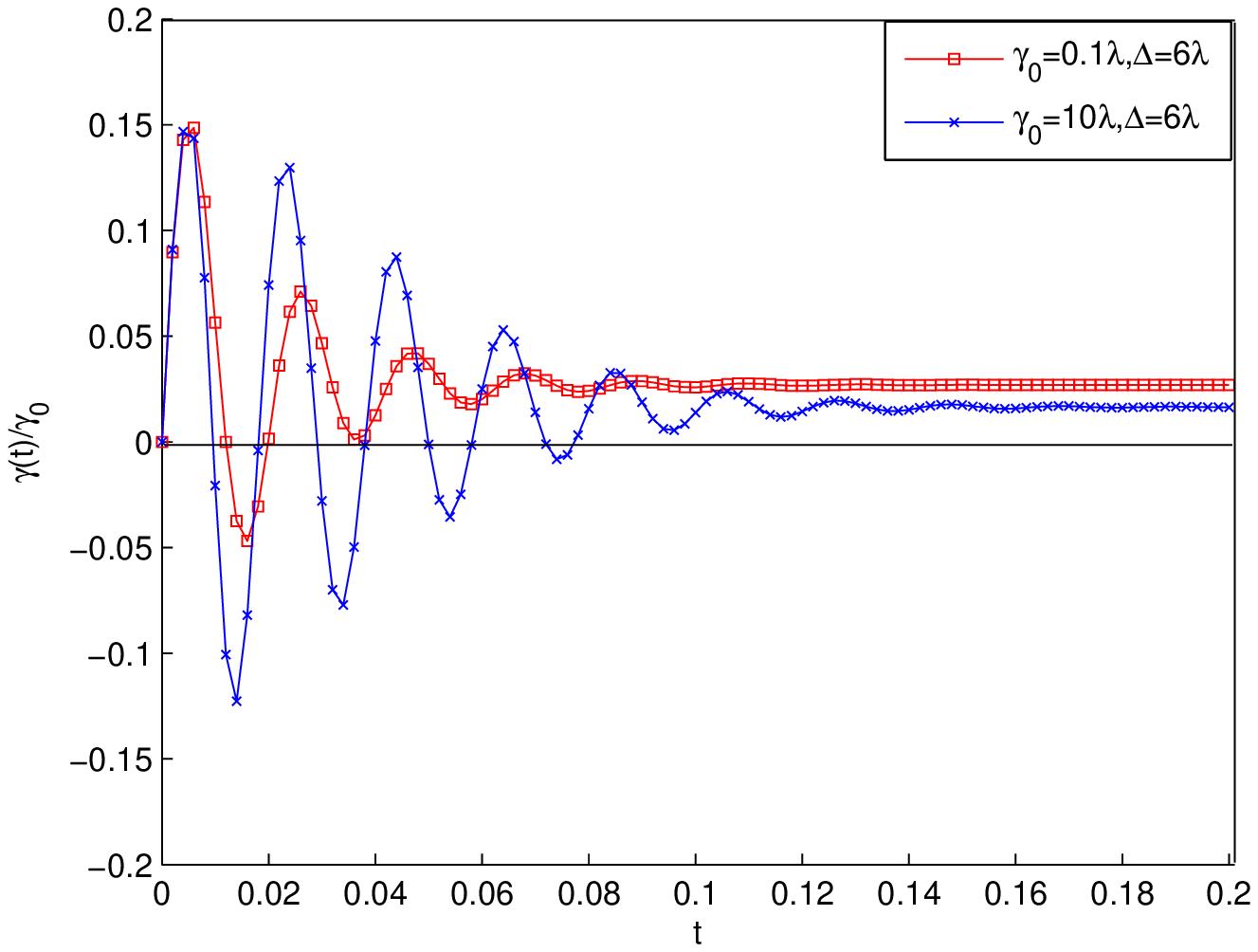}\\
  \caption{$(a)$ : $\tau_{QSL}/\tau_D$ is plotted as a function of the evolved time $\tau$ in the off-resonance case.
  The red dashed line represents the result of the weak-coupling regime, $\gamma_0=0.1\lambda, \; \Delta=6\lambda$.
  The blue solid line represents the result of the strong-coupling regime, $\gamma_0=10\lambda, \; \Delta=6\lambda$. $(b)$ : the off-resonance decay rate $\gamma(t)/\gamma_0$ changes with time. The red squares denote the condition of $\gamma_0=0.1\lambda, \; \Delta=6\lambda$. The blue stars denote that of $\gamma_0=10\lambda, \; \Delta=6\lambda$.}\label{6}
\end{figure}

\end{document}